\documentstyle[12pt]{article}
\input epsf.tex

\newenvironment{frontpage}{%
	\setcounter{page}{0}
	\pagestyle{empty}}
	{\newpage
	\setcounter{page}{1}}

\newcommand{\preprint}[1]{%
	{\rightline{#1}}}

\renewcommand{\title}[1]{%
	{\begin{center}
	\Large\bf #1
	\end{center}}
	\vskip .3in}

\renewcommand{\author}[1]{%
	{\begin{center}
	#1
	\end{center}}}

\newcommand{\institution}[1]{\vspace{-1.7em}\vspace{0pt}
	{\begin{center}
	\it #1
	\end{center}}}

\renewcommand{\abstract}[1]{%
	\begin{center}%
	{\vspace{1em}\vspace{0pt}\bf Abstract}%
	\end{center}%
	\noindent #1}

\renewcommand{\date}[1]{%
	\begin{center}%
	#1%
	\end{center}}

\makeatletter

\renewcommand{\section}{\@startsection{section}{1}{\z@}%
    {-3.5ex plus -1ex minus -.2ex}%
    {2.3ex plus.2ex}%
    {\centering\normalsize\bf}}

\renewcommand{\subsection}{\@startsection{subsection}{2}{\z@}%
	{-3.25ex plus -1ex minus -.2ex}%
	{1.5ex plus .2ex}%
	{\centering\normalsize\it}}

\makeatother

\newenvironment{references}{%
	\newpage
	\begin{thebibliography}{99}%
	\frenchspacing}
	{\end{thebibliography}}

\makeatletter
	\@addtoreset{equation}{section}%
\makeatother

\newcommand{\eqn}[1]{\label{eq:#1}}

\newcommand{\refeq}[1]{(\ref{eq:#1})}

\newcommand{\Eq}{Eq.~\refeq}

\newcommand{\beq}{\begin{eqnarray}}
\newcommand{\eeq}{\end{eqnarray}}

\newcommand{\rep}{representation}

\newcommand{\naive}{{na\"\i ve}}
\newcommand{\naively}{{na\"\i vely}}

\def\endignore{}
\def\ignore #1\endignore{}

\newcommand{\tr}{\mathop{\rm tr}}

\newcommand{\twi}{\widetilde}
\newcommand{\mybar}[1]%
	{\kern 0.8pt\overline{\kern -0.8pt#1\kern -0.8pt}\kern 0.8pt}
\newcommand{\sla}[1]%
	{\raise.15ex\hbox{$/$}\kern-.57em #1}
\newcommand{\roughly}[1]%
	{\mathrel{\raise.3ex\hbox{$#1$\kern-.75em\lower1ex\hbox{$\sim$}}}}

\newcommand{\drawsquare}[2]{\hbox{%
\rule{#2pt}{#1pt}\hskip-#2pt
\rule{#1pt}{#2pt}\hskip-#1pt
\rule[#1pt]{#1pt}{#2pt}}\rule[#1pt]{#2pt}{#2pt}\hskip-#2pt
\rule{#2pt}{#1pt}}

\newcommand{\Yfund}{\raisebox{-.5pt}{\drawsquare{6.5}{0.4}}}
\newcommand{\Ysymm}{\raisebox{-.5pt}{\drawsquare{6.5}{0.4}}\hskip-0.4pt%
        \raisebox{-.5pt}{\drawsquare{6.5}{0.4}}}
\newcommand{\Yasymm}{\raisebox{-3.5pt}{\drawsquare{6.5}{0.4}}\hskip-6.9pt%
        \raisebox{3pt}{\drawsquare{6.5}{0.4}}}

\newcommand{\avg}[1]{\langle #1 \rangle}

\newcommand{\nop}[1]{:\kern-.3em#1\kern-.3em:}

\newcommand{\Group}[2]{{\hbox{{\sl #1}($#2$)}}}
\newcommand{\U}[1]{\Group{U\kern0.05em}{#1}}
\newcommand{\SU}[1]{\Group{SU\kern0.1em}{#1}}
\newcommand{\SL}[1]{\Group{SL\kern0.05em}{#1}}
\newcommand{\Sp}[1]{\Group{Sp\kern0.05em}{#1}}
\newcommand{\SO}[1]{\Group{SO\kern0.1em}{#1}}

\newcommand{\hepth}[1]{{\tt hep-th/#1}}

\newcommand{\jref}[4]{{\it #1} {\bf #2}, #3 (#4)}

\newcommand{\CMP}[3]{\jref{Comm.\ Math.\ Phys}{#1}{#2}{#3}}

\newcommand{\NPB}[3]{\jref{Nucl.\ Phys.}{B#1}{#2}{#3}}

\newcommand{\PLB}[3]{\jref{Phys.\ Lett.}{#1B}{#2}{#3}}

\newcommand{\PRD}[3]{\jref{Phys.\ Rev.}{D#1}{#2}{#3}}

\newcommand{\ga}{{\gamma}}

\newcommand{\de}{{\delta}}

\newcommand{\ep}{{\epsilon}}

\newcommand{\la}{{\lambda}}

\newcommand{\si}{{\sigma}}

\def\twi{\widetilde}
\def\dtwi#1{\skew1\widetilde{\widetilde{#1}}}

\def\vbr{\vphantom{\sqrt{F_e^i}}}

\newcommand{\susc}{supersymmetric}
\newcommand{\spot}{superpotential}

\begin{document}

\begin{frontpage}

\preprint{UMD-PP-96-71}
\preprint{BUHEP-96-6}

\title{A Sequence of Duals\\
for \Sp{2N} Supersymmetric Gauge Theories\\
with Adjoint Matter}

\author{Markus A. Luty}

\institution{Department of Physics\\
University of Maryland\\
College Park, Maryland 20742}

\author{Martin Schmaltz\ \ and\ \ John Terning}

\institution{Department of Physics\\
Boston University\\
590 Commonwealth Avenue\\
Boston, Massachusetts 02215}

\date{March, 1996}

\abstract{
We consider supersymmetric \Sp{2N} gauge theories with $F$ matter fields in
the defining representation, one matter field in the adjoint representation,
and no superpotential.
We construct a sequence of dual descriptions of this theory using the
dualities of Seiberg combined with the ``deconfinement'' method introduced
by Berkooz.
Our duals hint at a new non-perturbative phenomenon that seems to be
taking place at asymptotically low energies in these theories:
for small $F$ some of the degrees of freedom form massless, non-interacting
bound states while the theory remains in an interacting non-Abelian Coulomb
phase.
This phenomenon is the result of strong coupling gauge dynamics in the
original description, but has a simple classical origin in the dual
descriptions.
The methods used for constructing these duals can be generalized to any
model involving arbitrary 2-index tensor representations of \Sp{2N},
\SO{N}, or \SU{N} groups.
}

\end{frontpage}

\section{Introduction}
Recently there has been considerable progress in understanding
non-perturbative effects in \susc\ gauge theories (see for example
Refs.~\cite{Holomo,Seib,Seib2,IntSeib}).
In particular, Seiberg \cite{Seib2} has argued convincingly that the
low-energy dynamics of supersymmetric \SU{N} QCD can be described by a dual
gauge theory with different gauge group and matter content.
(The \SO{N} and \Sp{2N} cases were worked out in detail in
Refs.~\cite{IntSeibSO} and \cite{IntPoul}, respectively.)
Dual descriptions have since been discovered for a wide range of theories;
see for example Refs.~\cite{IntLeighStrass,Berkooz,Pouliot}.

One theory that has attracted considerable attention recently is the
model with gauge group \SU{N}, \SO{N}, or \Sp{2N}, containing $F$ vector-like
``flavors'' of matter fields in the defining \rep\ and one matter field,
$A$, in the adjoint \rep\ \cite{Kut,KutSchw,LeighStrass,KutSchwSeib}.
Kutasov and Schwimmer \cite{Kut,KutSchw} have constructed dual descriptions
for the \SU{N} theory with the addition of a \spot\ of the form
\beq
\eqn{treesup}
W = \tr(A^{k + 1}).
\eeq

The \SO{N} and \Sp{2N} analogs of this model were worked out in
Ref.~\cite{LeighStrass} (see also Ref.~\cite{IntSpSO}).
It was found that the size of the dual gauge group depends on $k$,
and becomes infinitely large as $k \to \infty$.
More recently, Kutasov, Schwimmer, and Seiberg \cite{KutSchwSeib}
have obtained impressive detailed
evidence for the validity of the duality presented in
Refs.~\cite{Kut,KutSchw}.

The dynamics of the theory without a \spot\ is still not well understood,
although there are some hints coming from analyzing the theory with various
\spot s added.
Using the dual descriptions of Refs.~\cite{Kut,KutSchw,KutSchwSeib} it can
be shown that in the presence of the \spot\ \Eq{treesup}, the theory is at
an interacting superconformal fixed point for $2 N / (2 k - 1) < F < 2 N$.
Taking the limit $k \to \infty$ (which makes the superpotential arbitrarily
flat at the origin) one can argue that the low-energy dynamics of the
theory without a \spot\ is described by an interacting superconformal fixed
point for $0 < F < 2 N$ \cite{KutSchwSeib}.
This suggestion is in accord with the results of Ref.~\cite{Elitzur}, which
studied the theory with the addition of a superpotential
\beq
W = \la \hat{Q} A Q.
\eeq
where $Q$ and $\hat{Q}$ are the fundamental and antifundamental matter fields,
respectively.
For finite $\la$ the theory is smoothly connected to $N = 2$
\susc\ QCD, which is known to be in an Abelian Coulomb phase
\cite{SeibergWitten,Elitzur}.
When the limit $\la \to 0$ is taken, it is found that singularities
appear in the low-energy effective action \cite{Elitzur}, as expected if the
theory is entering a non-Abelian Coulomb phase, and the requisite gauge
bosons are becoming massless.
In this paper, we further investigate the conjecture that the theory without
a superpotential is at an interacting superconformal fixed point for all
$0 < F < 2 N$.

We will study the \Sp{2N} version of this theory.
Its properties are expected to be very similar to the \SU{N} case, but it
is easier to analyze because the invariants of \Sp{2N} are simpler.
Assuming that the theory is at a non-trivial superconformal fixed point,
we show that some massless degrees of freedom must decouple from the
superconformal fixed point theory for $F < F_0$, where
$F_0 \ge \frac{1}{2}(N + 1)$.
We describe the various possibilities for which operators decouple in the
infrared.

We then construct a series of dual descriptions that suggest that
gauge invariant operators of the form $Q A^k Q$, $k = 0,1,2,\ldots$
sequentially decouple as $F$ is reduced.
The dual descriptions are constructed by generalizing the ``deconfinement''
method introduced by Berkooz \cite{Berkooz,Pouliot}.
The first dual description is a theory with gauge group
${\sl Sp} \times {\sl SO}$, with matter fields in the fundamental and adjoint
\rep s, and a \spot.
We then iterate the process, applying the deconfinement
method to {\sl SO} and {\sl Sp} groups to obtain more complicated dual
descriptions.%
\footnote{A similar series of duals for an {\sl SU} theory with
an antisymmetric tensor was noted in \cite{Pouliot}.}

This paper is organized as follows.
In Section 2 we introduce the theory under investigation and consider some
of the possible scenarios for the infrared physics.
In Section 3 we describe the construction of the first dual, and in Section
4 we iterate the construction to obtain additional duals that we then use to
speculate about the infrared spectrum.
In the appendices, we perform some consistency checks on these dual
descriptions and show how the deconfinement method can be generalized to
arbitrary 2-index \rep s.

\section{The Model}
The theory we wish to study has gauge group $\Sp{2N}$ with $2F$ matter fields
$Q$ in the defining \rep, and one matter field $A$ in the adjoint
(symmetric second rank tensor) \rep.
It has the anomaly-free global symmetry $\SU{2F} \times \U{1} \times
\U{1}_R$.
The field content (with global charges) is given in table 1.

\begin{table}[htbp]
\centering
\begin{tabular}{|c||c||c|c|c|}\hline
field & \Sp{2N} & \SU{2F} & \U{1} & $\U{1}_R$
\\\hline \hline
$Q$ & \Yfund & \Yfund & ${{N+1}\over{F}}$ & $1\vbr$ \\  \hline
$A$ & \Ysymm & {\bf 1} & $-1$ & $0\vbr$ \\
\hline
\end{tabular}
\label{Sp}
\parbox{4in}{\caption{Field content of the theory.}}
\end{table}

The moduli space can be parameterized by the holomorphic gauge-invariant
polynomials in the matter fields (the ``chiral ring'').%
\footnote{This connection has been known for some time;
for a proof, see Ref.~\cite{poly}.}
In the present case, these are generated by
\beq
T_k & \equiv & \tr A^{2k}, \quad k = 1, 2, \ldots
\\
M_k & \equiv & Q A^k Q, \quad k = 0, 1, \ldots
\eeq
In Ref.~\cite{LeighStrass} (following Refs.~\cite{Kut,KutSchw}) this
theory is studied with the addition of a \spot
\beq
W = \tr A^{2 k},
\eeq
and different dual descriptions are constructed for each value of $k \ge 1$.
The dual gauge group is \Sp{2\twi{N}}, where
$\twi{N} = (2 k + 1) F - (N + 2)$.

We will consider the theory with no \spot.
We will often take advantage of the simplifications of the large $N$ limit
(with $N/F$ held fixed), but we believe that most of these results remain
valid for $N \sim 1$.
The exact $\beta$-function of the theory satisfies \cite{betaref}
\beq
\eqn{russbeta}
\beta \propto 2 (N + 1) - F (1 - \ga_{QQ}) + (N+1) \ga_{AA},
\eeq
where $\ga_{QQ}$ and $\ga_{AA}$ are the anomalous dimensions of the operators
$QQ$
and $\tr A^2$, respectively.
The theory is infrared free for $F \ge 2 (N + 1)$.
For $F$ just below $2 (N + 1)$ the one-loop coefficient
of the $\beta$-function is small and positive while the two-loop coefficient
is negative of order $N^2$.
Arguments due to Banks and Zaks \cite{BanksZaks} show that the gauge coupling
has a non-trivial perturbative fixed point in the infrared (at least in the
large-$N$ limit).
At the fixed point we can calculate anomalous dimensions in perturbation
theory:
\begin{eqnarray}
\gamma_{AA} & =&  -{\alpha_* \over \pi} (N+1)  + O(\alpha_*^2)  \nonumber \\
\gamma_{QQ} & =&  -{\alpha_* \over 2 \pi} (N+\frac12)  + O(\alpha_*^2)
\end{eqnarray}
where $\alpha_*$ is the gauge coupling at the fixed point, given by
\beq
{\alpha_* \over  \pi}(N+1) = {\epsilon \over 2} + O(\epsilon^2),
\qquad \epsilon \equiv 2 - {F \over N+1}.
\eeq
The superconformal algebra in the infrared includes an anomaly-free
$R$ symmetry whose charges,  $R_{\rm SC}$, are related to the dimensions of
chiral
operators ${\cal O}$ by \cite{sconf}
\beq
\dim({\cal O}) = {3 \over 2} R_{\rm SC}({\cal O}).
\eeq
The $R_{\rm SC}$ charges must be a linear combinations of the anomaly-free
$U(1)$ symmetries of the ultraviolet:
\beq
\eqn{Rsuperconf}
R_{\rm SC} = R - b U
\eeq
where $R$ and $U$ denote charges under the $U(1)_R$ and $U(1)$,
respectively.
The coefficient $b$ is a function of $F$ and $N$ that we would like to
determine, since $b$ determines the scaling dimensions of the operators in
the superconformal algebra.
Near the Banks--Zaks fixed point
\beq
b = \frac{2}{3} - \frac{\ep}{6} + O(\ep^2).
\eeq
As we decrease $F$ away from $2(N + 1)$ the gauge coupling at the fixed point
increases; eventually perturbation theory breaks down and we
do not know how to determine $b$.
However, assuming that the theory is at a conformal fixed point in the
infrared we can find limits on $b$ by using the fact that the dimensions
of gauge invariant chiral operators satisfy the unitarity bound \cite{sconf}
$\dim({\cal O}) \ge 1$.
The bound is saturated for free fields.
In Fig.~1 we show the values of $b$ (as a function of $F$) for which the
operators in the chiral ring saturate the bound:
\beq
\dim(T_k) = 3 k b = 1,
\qquad
\dim(M_k) = 3  - 3 b \left( \frac{N + 1}{F} - \frac{k}{2} \right) = 1.
\eeq

Assuming that the gauge coupling is at a fixed point in the infrared for
$0 < F < 2 (N + 1)$, for large $N$ we can imagine three qualitatively
different scenarios (see Fig.~1):

\begin{figure}[htbp]
\centering
\epsfxsize=5.0in\epsfbox{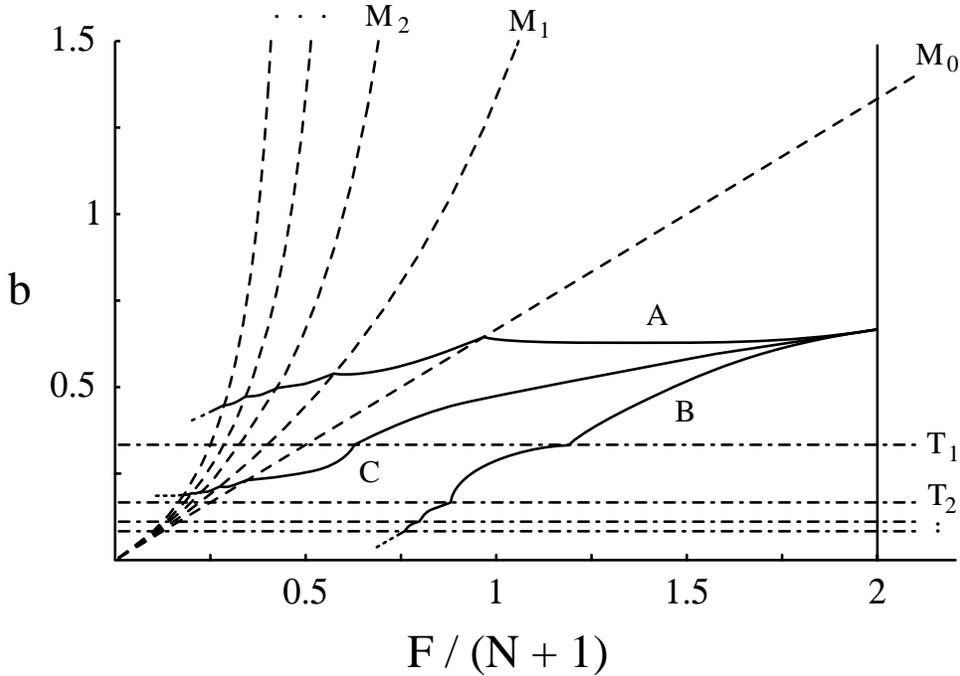}
\smallskip
\parbox{4in}{\caption{The scenarios A, B, and C for the behavior of the
coefficient $b$ that determines the superconformal $R_{\rm SC}$ charges as a
function of $F / (N + 1)$, for large, fixed $N$.
The curves meet at $F = 2(N + 1)$, where the position and slope can
be calculated in perturbation theory.
For $F$ near zero it presumably does not make sense to plot $b$ as
a continuous function.
The lines labeled $M_0, M_1, \ldots$ and $T_1, T_2, \ldots$ indicate the
region where the corresponding operators have scaling dimensions of a free
field.}}
\end{figure}

\begin{itemize}
\item[A.]
$b$ successively crosses the lines corresponding to $M_0, M_1, \ldots$ as
$F$ is reduced.
At the value $F = F_0$ where $b$ crosses the $M_0$ line, the dimension of
$M_0$ as calculated using \Eq{Rsuperconf} violates the unitarity bound.
This implies that $M_0$ is a free field (with $\dim(M_0) = 1$) for
$F \le F_0$.
The theory can still be at an interacting superconformal fixed point for
$F < F_0$, since there is an accidental $R_0$ symmetry in the infrared under
which only the free field transforms.
This $R_0$ symmetry redefines the superconformal $R$ charge of $M_0$ and
allows the dimension of $M_0$ to stay one.
The other fields do not transform under the accidental $R_0$ symmetry,
and therefore the dimensions of all other operators in the theory
are still given by \Eq{Rsuperconf}.
We will therefore assume that the the theory remains at a superconformal
fixed point.
Since the number of degrees of freedom is of order $N^2$, and we are changing
the number of degrees of freedom by order 1 in crossing the $M_0$ line, it is
plausible that $b$ is continuous across the line as shown in Fig.~1.
Analogous effects occur as $b$ crosses the lines corresponding to the
operators $M_1, M_2, \ldots$.

\item[B.]
$b$ crosses the lines corresponding to $T_1, T_2, \ldots$ as $F$ is reduced.
In this scenario the operators $T_k$ sequentially decouple in a similar
fashion as in scenario A above.

\item[C.]
$b$ crosses lines corresponding to both types of operators $M_k$ and $T_k$.
\end{itemize}
Without further information we cannot decide which of these scenarios is
correct.
In the next sections we will construct dual descriptions that we will use to
argue in favor of the first scenario.
The operators  $M_k$ appear as fundamental fields in the duals and there
appear to be values of $F$ and $N$ where they are free fields.

\section{Construction of the First Dual}
We can find another \susc\ gauge theory that has the same low-energy
dynamics using the ``deconfinement'' method of Berkooz
\cite{Berkooz,Pouliot}.%
(In fact this method is quite general and can be used to write a
dual description of almost any \susc\ gauge theory.
See Appendix B.)
The idea is to replace the adjoint by a composite ``meson'' operator of a
strongly-interacting \SO{N'} group:
\beq
A^{ab} \to x^{aa'} x^{ba'},
\eeq
where $a, b$ are \Sp{2N} indices and $a'$ is an \SO{N'} index.
$N'$ can be chosen so that \SO{N'} confines without chiral symmetry breaking
\cite{Seib,IntSeibSO}.%
\footnote{The branch of the \SO{N'} theory that
generates a dynamical superpotential is eliminated because it has no vacuum.}
The problem with a straightforward application of this idea is that the
$\Sp{2N} \times \SO{N'}$ theory has one less anomaly-free $\U{1}$ symmetry,
since there is an additional constraint that the $\U{1}$ symmetries have no
\SO{N'} anomaly.
This problem can be circumvented by introducing additional fields that
transform under \SO{N'} and adding terms to the \spot\ in the deconfined
description.
The matter content of the model that accomplishes this is displayed in
Table \ref{SpSO}.
The \spot\ in the deconfined description is
\beq
W = x_1 p_1 p_2 + p_1 p_1 p_3.
\eeq
(We have set the coefficients of the superpotential to $+1$ by rescaling
the fields.)
The purpose of the \spot\ is to give masses to the the unwanted ``meson''
states $(x_1 p_1)$ and $(p_1 p_1)$ that appear when the \SO{N'} group
confines.%
\footnote{In generalizations of this deconfinement method to other groups,
the confining gauge group generates a dynamical \spot\ for the composite
fields.
In these cases, the analogs of the \spot\ terms discussed above
also serve to eliminate this \spot\ via their equations of motion.
See Appendix B.}
With this matter content, we take $N' = 2N + 5$.
Note that the $\U{1}$ charges are uniquely determined by the following
constraints:
they agree with the $\U{1}$ charges in the original theory,
they are anomaly free,
and the \spot\ is invariant.
We explicitly check that all anomalies of the original theory match those of
the dual in appendix A.

\begin{table}[htbp]
\label{SpSO}
\centering
\begin{tabular}{|c||c|c||c|c|c|}\hline
field & \Sp{2N} & \SO{2N + 5} & \SU{2F} & \U{1} & $\U{1}_R$
\\
\hline\hline
$Q$ & \Yfund & {\bf 1} & \Yfund & ${{N+1}\over{F}}$ & $1\vbr$ \\ \hline
$x_1$ & \Yfund & \Yfund & $ {\bf 1}$ & $-{{1}\over{2}}$ & $0\vbr$ \\ \hline
$p_1$ & {\bf 1} & \Yfund & {\bf 1} & $N$ & $-2\vbr$ \\ \hline
$p_2$ & \Yfund & {\bf 1} & {\bf 1} & $\frac{1}{2} - N$ & $4\vbr$ \\ \hline
$p_3$ & {\bf 1} & {\bf 1} & {\bf 1} & $-2N$ & $6\vbr$\\
\hline
\end{tabular}
\parbox{4in}{\caption{Field content of the first ``deconfined" theory.}}
\end{table}

We can now use the known dual description of \Sp{2N} gauge theory with
fundamentals \cite{Seib2,IntPoul} to write a dual description of this theory
in terms of a gauge theory with gauge group
$\Sp{2F + 2} \times \SO{2N + 5}$.
The field content is given in Table 3;
the dual has \spot
\beq
W & = & M_0 \twi{Q} \twi{Q} + A_1 \twi{x}_1 \twi{x}_1 + m_1 \twi{Q}
\twi{x}_1 + m_2 \twi{Q} \twi{p}_2 + m_3 \twi{x}_1 \twi{p}_2
\nonumber\\
&& \quad +\, m_3 p_1 + p_1 p_1 p_3.
\eeq
We can integrate out the massive fields $m_3$ and $p_1$, leaving the \spot\
(after field rescaling)
\beq
\label{firstdualW}
W & = & M_0 \twi{Q} \twi{Q} + A_1 \twi{x}_1 \twi{x}_1 + m_1 \twi{Q}
\twi{x}_1 + m_2 \twi{Q} \twi{p}_2
+ (\twi{x}_1 \twi{p}_2) (\twi{x}_1 \twi{p}_2) p_3.
\eeq
The anomaly matching is guaranteed to work by the anomaly matching of the
{\sl Sp} duality used in the construction.
The gauge-invariant chiral operators of the original theory map into the
dual description as follows:
\beq
\tr A^{2k} & \to & \tr A_1^{2k}, \quad k = 1, 2 \ldots
\nonumber\\
Q Q & \to & M_0,
\\
Q A^k Q & \to & m_1 A_1^{k - 1} m_1, \quad k = 1, 2, \ldots\,.
\nonumber
\eeq
Note that the composite operator $M_0 \equiv QQ$ of the original theory is a
fundamental field in the dual description.
This is similar to Seiberg's dual description of supersymmetric QCD
\cite{Seib2}.
However, the theory we are considering here has a much more complicated
structure;
there are many operators in the chiral ring that do not map onto elementary
fields in the first dual.
We will see in the following section that the operators
$M_k \equiv Q A^{k} Q$ map onto fundamental fields in the $n$-th dual for
$n > k$, while operators of the form $T_k \equiv \tr A^{2k}$ never appear as
fundamental fields in our duals.

\begin{table}[htbp]
\label{SpSODual}
\centering
\begin{tabular}{|c||c|c||c|c|c|}\hline
field &$\Sp{2F + 2}$ &$\SO{2N + 5}$& $\SU{2F}$ & $\U{1}$ & $\U{1}_R$
\\\hline \hline
$\twi{Q}$ &  \Yfund&${\bf  1}$&$\overline{\Yfund}$&$
-{{N+1}\over{F}}$&$0   \vbr$ \\ \hline
 $M_0$ & $ {\bf 1}$&${\bf  1}$&\Yasymm&$2
{{N+1}\over{F}}$&$2   \vbr$ \\ \hline
$\twi{x}_1$ &  \Yfund& \Yfund&$ {\bf 1}$&${{1}\over{2}}$&$ 1
  \vbr$ \\ \hline
$A_1$ & $ {\bf 1}$& $ \Yasymm$&{\bf 1}&$-1$&$0
 \vbr$ \\ \hline
$m_1$ & $ {\bf 1}$& \Yfund& \Yfund&$
{{N+1}\over{F}}-{{1}\over{2}}$&$1   \vbr$ \\ \hline
$m_2$ & {\bf 1}&$ {\bf 1}$& \Yfund&
${{N+1}\over{F}}+{{1}\over{2}}- N$&$5   \vbr$ \\ \hline
$\twi{p}_2$ & \Yfund&$ {\bf 1}$&$ {\bf 1}$&$-{{1}\over{2}}+N$&$-3
  \vbr$ \\ \hline
$p_3$ & $ {\bf 1}$&$ {\bf 1}$&$ {\bf 1}$&$-2 N$&$ 6\vbr$
\\ \hline
$p_1$ & $ {\bf  1}$& \Yfund&$ {\bf 1}$&$N$&$-2   \vbr$ \\
\hline
 $m_3$& $ {\bf 1}$& \Yfund&$ {\bf 1}$&$- N $&$4\vbr$\\
\hline
\end{tabular}
\parbox{4in}{\caption{Field content of the first dual description.}}
\end{table}

What dynamical information can we obtain from this dual?
One might have hoped that our dual descriptions would be free in the
infrared for some range of $N$ and $F$, so that our dual gives a
weakly-coupled description of the low-energy physics.
(This happens in the dual description of supersymmetric QCD for
$N + 1 < F < \frac{3}{2}N$.)
At one loop, the {\sl SO} group is infrared free for $F \ge N + 1$ and the
{\sl Sp} group is infrared free for $F \le \frac{1}{2} (N - 3)$.
Therefore, this dual description is never completely free in the infrared.
(A similar situation holds in all of our duals.)
This is not surprising, since we expect that the theory is at an interacting
superconformal fixed point, and such a theory cannot have a dual description
that is free in the infrared.

In the range $N + 1 \le F < 2N$ the {\sl SO} gauge group is infrared free
(at one loop), and the {\sl Sp} gauge group has the right number of colors
and flavors (again at one loop) to be in the ``conformal window'' where
there is an interacting superconformal fixed point \cite{Seib2,IntPoul}.
The one-loop calculation amounts to neglecting the contributions of the
anomalous dimensions to the $\beta$ function in \Eq{russbeta}; this gives
qualitatively wrong results when the anomalous dimensions are large due to
relevant interactions in the superpotential or strong gauge interactions.
An example of such a situation is our dual for $F\roughly< 2N$.
The (incorrect) one-loop calculation suggests that the {\sl SO} interactions
can be ignored in the infrared. Ignoring {\sl SO} we would find that the
{\sl Sp} gauge
group is near the end of its ``conformal window" and the anomalous dimension
of the {\sl Sp}-gauge invariant operator $\twi{x}_1 \twi{x}_1$ would be near
1 in analogy to supersymmetric QCD.
Then the relevant term in the superpotential $A_1 \twi{x}_1 \twi{x}_1$ would
force the dimension of $A_1$ to be near 2.
But by the operator map, $\tr A_1^{2k}$ corresponds to the operator $\tr
A^{2k}$ in the original description, and the dimension of $A$ is close
to 1 at the Banks--Zaks fixed point, thus contradicting our \naive\
interpretation of the dual.

This example illustrates that a one-loop calculation of $\beta$ functions
is not reliable.
The reason is that relevant superpotential couplings and strong gauge groups
contribute large anomalous dimensions to the fields of the theory which
cannot be ignored when calculating $\beta$ functions.
The gauge group that was \naively\ believed to be free may even
be rendered relevant in the infrared via the anomalous dimensions.
This may occur whenever there are fields transforming under both groups, or
when there are relevant superpotential couplings.
Therefore, one must be very careful in trying to draw physical conclusions
from a dual description that is not weakly coupled.

Another obstacle to extracting low-energy physics from this dual is that in
the deconfined theory, we introduced massive degrees of freedom in order to
cancel anomalies.
Once we pass from the deconfined description to the dual description, the
fact that these degrees of freedom are irrelevant in the infrared is no
longer evident.

While keeping these points in mind, we will nonetheless use this dual and
its generalizations below to argue that the operators $M_k$ are free fields
for sufficiently small $F$.
In Appendix A we will also perform some consistency checks on this dual
description.
These help convince us that the dual description is correct, but by
themselves they do not give us dynamical information that we do not already
know from the original description of the theory.

\section{More Dual Descriptions}
We can obtain additional dual descriptions by applying the  deconfinement
method again, this time to the adjoint of the {\sl SO} group in the first
dual.
To this end, we introduce a confining {\sl Sp} group that forms a composite
meson with the same quantum numbers as the {\sl SO} adjoint $A_1$.
(This is precisely the version of deconfinement discussed in
Ref.~\cite{Pouliot}.)
The field content is given in Table 4, and the superpotential is
\beq
W = M_0 \twi{Q} \twi{Q} + (x_2 x_2) \twi{x}_1 \twi{x}_1 + m_1 \twi{Q}
\twi{x}_1 + m_2 \twi{Q} \twi{p}_2
+ (\twi{x}_1 \twi{p}_2) (\twi{x}_1 \twi{p}_2) p_3 + x_2 r_1 r_2.
\eeq

\begin{table}[htbp]
\label{SpSO5Sp}
\centering
\begin{tabular}{|c||c|c|c||c|c|c|}\hline
field &$\Sp{2F + 2}$ &$\SO{2N + 5}$&$\Sp{2N + 2}$& $\SU{2F}$ & $\U{1}$ &
$\U{1}_R$
\\\hline \hline
 $\twi{Q}$ &  \Yfund&${\bf  1}$&$ {\bf 1}$&$ \overline{\Yfund}$&$
-{{N+1}\over{F}}$&$0   \vbr$ \\ \hline
 $M_0$ & $ {\bf 1}$&${\bf  1}$&${\bf  1}$&\Yasymm&$2
{{N+1}\over{F}}$&$2   \vbr$ \\ \hline
$\twi{x}_1$ &  \Yfund& \Yfund&$ {\bf 1}$&${\bf  1}$&${{1}\over{2}}$&$ 1
  \vbr$ \\ \hline
$x_2$ & $ {\bf 1}$& \Yfund & \Yfund &{\bf 1}&$-\frac{1}{2}$&$0
 \vbr$ \\ \hline
 $r_1$ & $ {\bf 1}$&$ {\bf 1}$& \Yfund&$ {\bf
1}$&$N+{{5}\over{2}}$&$2   \vbr$ \\ \hline
 $r_2$ & $ {\bf 1}$& \Yfund&$ {\bf 1}$&{\bf 1}&$-N-2$&$0
  \vbr$ \\ \hline
$m_1$ & $ {\bf 1}$& \Yfund&$ {\bf 1}$& \Yfund&$
{{N+1}\over{F}}-{{1}\over{2}}$&$1   \vbr$ \\ \hline
$m_2$ & {\bf 1}&$ {\bf 1}$&$ {\bf 1}$& \Yfund&
${{N+1}\over{F}}+{{1}\over{2}}- N$&$5   \vbr$ \\ \hline
$\twi{p}_2$ & \Yfund&$ {\bf 1}$&$ {\bf 1}$&${\bf  1}$&$-{{1}\over{2}}+N$&$-3
  \vbr$ \\ \hline
$p_3$ & $ {\bf 1}$&$ {\bf 1}$&$ {\bf 1}$&${\bf  1}$&$-2N$&$6
  \vbr$ \\ \hline
\end{tabular}
\parbox{4in}{\caption{Field content of the second ``deconfined''
description.}}
\end{table}

We can now use the known dual description of {\sl SO} gauge theory
with fundamentals \cite{Seib2,IntSeibSO} to write a dual description of
this theory in terms of a theory with gauge group
$\Sp{2F + 2} \times \SO{4F + 4} \times \Sp{2N + 2}$.
Some of the fields are massive and can be integrated out.
The field content of the resulting theory is given in
Table 5, and the superpotential is
\beq
W & = & M_0 (\dtwi{x}_1 \twi{m}_1) (\dtwi{x}_1 \twi{m}_1)
+ (\dtwi{x}_1 \twi{x}_2) (\dtwi{x}_1 \twi{x}_2)
+ m_2 \twi{p}_2 (\dtwi{x}_1 \twi{m}_1)
+ n_1 \twi{p}_2^2 p_3
\nonumber\\
&& \quad +\, n_1 \dtwi{x}_1 \dtwi{x}_1
+ A_2 \twi{x}_2 \twi{x}_2 + M_1 \twi{m}_1 \twi{m}_1
+ n_3 \twi{r}_2 \twi{r}_2
\nonumber\\
&& \quad +\, n_2 \twi{x}_2 \twi{m}_1 + n_4 \dtwi{x}_1 \twi{r}_2
+ n_5 \twi{m}_1 \twi{r}_2.
\eeq
The gauge-invariant chiral operators of the original theory map into the
second dual description as follows:
\beq
\begin{array}{clcll}
\tr A^{2k} & \to & \tr A_1^{2k} &\to & \tr A_2^{2k},
\quad k = 1, 2, \ldots
\\
Q Q & \to & M_0 & \to & M_0,
\\
Q A Q & \to & m_1 m_1 & \to & M_1,
\\
Q A^k Q & \to & m_1 A_1^{k - 1} m_1 & \to &  n_2 A_2^{k - 2} n_2,
\quad k = 2, 3, \ldots\,.
\end{array}
\eeq

\begin{table}[htbp]
\label{final}
\centering
\begin{tabular}{|c||c|c|c||c|c|c|}\hline
field &$\Sp{2F + 2}$ &$\SO{4F + 4}$&$\Sp{2N + 2}$& $\SU{2F}$ & $\U{1}$ &
$\U{1}_R$
\\\hline \hline
 $M_0$ & $ {\bf 1}$&${\bf  1}$&$ {\bf 1}$&\Yasymm&$2
{{N+1}\over{F}}$&$2   \vbr$ \\ \hline
$\dtwi{x}_1$ & \Yfund&\Yfund&$ {\bf 1}$&$ {\bf 1}$&$-{{1}\over{2}}$&$ 0
  \vbr$ \\ \hline
 $n_1$ & \Ysymm &${\bf  1}$&$ {\bf 1}$&{\bf 1}&$1$&$2
  \vbr$ \\ \hline
$\twi{x}_2$ &$ {\bf 1}$&\Yfund&\Yfund&$ {\bf
1}$&${{1}\over{2}}$&$1   \vbr$ \\ \hline
 $A_2$ & $ {\bf 1}$&$ {\bf 1}$& \Ysymm &$ {\bf 1}$&$
-1$&$0   \vbr$ \\ \hline
 $\twi{m}_1$ & $ {\bf 1}$&\Yfund&$ {\bf 1}$&$ \overline{\Yfund}$&$
{{1}\over{2}}-{{N+1}\over{F}}$&$0   \vbr$ \\ \hline
$M_1$ & $ {\bf 1}$&${\bf  1}$&$ {\bf 1}$& \Ysymm
&$2{{N+1}\over{F}}-1$&$2   \vbr$ \\ \hline
 $n_2$ &  {\bf 1}&$ {\bf 1}$&\Yfund&\Yfund&$
{{N+1}\over{F}}-1$&$1   \vbr$ \\ \hline
 $n_3$ & $ {\bf 1}$&$ {\bf 1}$&$ {\bf 1}$&$ {\bf 1}$&$-2 N -4$&$0
  \vbr$ \\ \hline
 $n_4$ & \Yfund&$ {\bf 1}$&$ {\bf 1}$&$ {\bf
1}$&$-N-{{3}\over{2}}$&$1   \vbr$ \\ \hline
 $n_5$ & $ {\bf 1}$&$ {\bf 1}$&$ {\bf 1}$&\Yfund&
$ {{N+1}\over{F}}-{{5}\over{2}}-N$&$1 \vbr$\\ \hline
 $m_2$ & $ {\bf 1}$&$ {\bf 1}$&$ {\bf 1}$& \Yfund&
${{N+1}\over{F}}+{{1}\over{2}}- N$&$5   \vbr$ \\ \hline
$\twi{p}_2$ & \Yfund&$ {\bf 1}$&$ {\bf 1}$&$ {\bf
1}$&$-{{1}\over{2}}+N$&$-3   \vbr$ \\ \hline
$p_3$ & $ {\bf 1}$&$ {\bf 1}$&$  {\bf 1}$&{\bf 1}&$-2 N$&$ 6
\vbr$ \\ \hline
 $\twi{r}_2$ &  {\bf 1}&\Yfund&$ {\bf 1}$&{\bf 1}&$N+2$&$1
  \vbr$ \\ \hline
\end{tabular}
\parbox{4in}{\caption{Field content of the second dual description
after integrating out massive fields.}}
\end{table}

As already stated in the previous section, both the composite operators
$M_0=QQ$ and $M_1 = QAQ$ of the original theory are fundamental
fields in this description.
Note that $M_0$ only interacts via the superpotential term
$M_0 (\dtwi{x}_1 \twi{m}_1) (\dtwi{x}_1 \twi{m}_1)$, which has
canonical dimension 5.
If we could identify a range of $F$ where this operator is irrelevant in the
infrared, we would have shown that $M_0$ is free in that
range.
For example, this would be the case if either one of the two gauge groups
\Sp{2F + 2} or \SO{4F + 4} were infrared free.
To see this, suppose that \SO{4F + 4} is infrared free.
(At one loop, we would naively conclude that this is the
case for $F \le \frac{1}{4} (N - 1)$.)
Then the superpotential term involving $M_0$ can be written as the product
of operators $M_0$, $\dtwi{x}_1 \dtwi{x}_1$, and $\twi{m}_1$ that are
gauge-invariant under all ``active'' gauge groups.
These operators must each have dimension at least 1, so the \spot\ term
involving $M_0$ has dimension at least 4, and $M_0$ is free in the
infrared.
This corresponds to the scenario A of Section 2.
Of course, the \spot\ and
the other gauge interactions do affect the range of $N$ and $F$ for which
the gauge groups are infrared free.
Nonetheless, because we know that {\em some} operators must become free,
we interpret this feature of our dual descriptions as suggesting that
scenario A is in fact correct.

Note that in the original description, the decoupling of the field $M_0$
from the superconformal algebra is a non-perturbative phenomenon.
In the dual description (provided we are interpreting it correctly), it
simply corresponds to the fact that $M_0$ couples only through a term in
the superpotential with high dimension, and the fact that $M_0$ is free
is a simple classical effect.
In this sense, our dual descriptions give a weakly-coupled description
of a strong coupling phenomenon in the original theory.

This feature also appears in the duals constructed by
Kutasov and Schwimmer \cite{Kut,KutSchw,KutSchwSeib} for the theory
with a superpotential \Eq{treesup}:
while the $T_k$ never appear as fundamental fields in their duals
the $M_k$ do, and they only couple through terms in the superpotential with
high
canonical dimensions. In the range of $F$ where the dual gauge group is free
these terms are irrelevant, and the $M_k$ become free fields.

One can continue constructing duals in this fashion.
We have constructed the third dual, and we summarize the matter content of the
original theory and the first three duals in Table 6.

\begin{table}[htbp]
\centering
\begin{tabular}{|c|c|c|}\hline
 dual & gauge group& matter
\\\hline \hline
 & \Sp{2N} & $2 F +{\bf A}$
\\\hline \hline
1 & \Sp{2 F + 2} & $ 2 F + 2 N + 6$ \\ \hline
 & \SO{2 N + 5} & $ 4 F +2 + {\bf A}$
\\\hline \hline
2 & \Sp{2 F + 2} & $ 4 F + 6 + {\bf A} $\\ \hline
& \SO{4F + 4} & $4 F + 2 N + 5$ \\ \hline
& \Sp{2 N + 2} & $ 6 F + 4 +{\bf A} $ \\ \hline
\hline
3 & \Sp{2 F + 2} & $ 4 F +6 + {\bf A} $\\ \hline
& \SO{4F + 4} & $8 F + 10 + {\bf A}$ \\ \hline
& \Sp{6 F + 6} & $ 6 F + 2 N + 12  $ \\ \hline
& \SO{2 N + 7} & $ 8 F + 6 + {\bf A}  $ \\ \hline
\hline
\end{tabular}
\label{summary}
\parbox{4in}{\caption{Matter content of all the first three duals.
${\bf A}$ indicates an adjoint.}}
\end{table}

A simple pattern emerges in these duals: in the $n$-th dual has $n+1$ gauge
groups with the following operator maps:
\beq
\begin{array}{cll}
\tr A^{2k} & \to & \tr A_n^{2k},
\quad k = 1, 2, \ldots
\\
Q A^k Q & \to &  M_k
\quad k = 1, \ldots, n - 1.
\end{array}
\eeq
The fields $M_k$ for $k < n - 1$ only interact via terms in the superpotential
that have large canonical dimensions.
It is therefore plausible that for sufficiently small $F$
such terms are irrelevant in the infrared.
(As above, one can show that this is the case provided that at least one of
the gauge groups is infrared free.)
As discussed above, this provides evidence that the operators $M_k$
successively become free as $F$ is reduced.

\section{Conclusions}
We have constructed the first three of an infinite sequence of dual
descriptions of an \Sp{2N} \susc\ gauge theory with matter fields
transforming as an adjoint and $F$ flavors, with no superpotential.
In the $n$-th dual description, the operators $M_k \equiv Q A^k Q$ appear
as fundamental fields for $k < n$, and the $M_k$ couple only through
superpotential interactions that have large canonical dimensions
and are likely to be irrelevant in the infrared.
This supports the scenario that the the operators $M_0, M_1, \ldots$
sequentially become free massless fields in the infrared as $F$ is reduced
from the asymptotic freedom limit $F = 2(N + 1)$.
In the original theory this picture arises from nonperturbative
quantum effects, while the dual descriptions give a simple classical
description of the same physics.

It would be very important to understand for which ranges of $N$ and $F$
the various gauge groups of our duals are weakly coupled, so that our
results could be put on firmer ground.
Unfortunately, this appears to be a difficult problem, partly due to the
interplay of the various different gauge groups, but also
because of the additional massive degrees of freedom that we
had to introduce in the ``deconfinement'' in order to match $U(1)$'s.

The extension of these results to {\sl SU} gauge theories is straightforward
using the ``deconfinement modules" discussed in Appendix B.

\section*{Acknowledgements}
We would like to thank A. Cohen, S. Chivukula, R.G. Leigh, L. Randall, and
W. Taylor IV for helpful discussions,
D. Kutasov for enlightening correspondence, and N. Seiberg for inspiration.
This work was supported in part by  the Department of Energy under
contracts \#DE-FG02-91ER40676 and  \#DE-AC02-76ER03069 and
by the National Science Foundation grant \#PHY89-04035 .

\appendix\newpage
\section{Consistency Checks}
In this appendix, we consider some consistency checks on the dual
descriptions constructed in the main text.

\subsection{Anomaly matching}
The dual descriptions were derived from known dualities, and so the
anomalies are guaranteed to match.
For completeness and to check our algebra, we explicitly computed the anomalies
in the original description and all of the the dual descriptions discussed
above,
with the following results:
\beq
SU(2 F)^3         &:& 2N
\nonumber\\
U(1)_R SU(2 F)^2  &:& 0
\nonumber\\
U(1)_R            &:& 0
\nonumber\\
U(1)_R^3          &:& 0
\nonumber\\
U(1) SU(2 F)^2    &:& \frac{2N(N + 1)}{F}
\\
U(1)              &:& N (2 N + 3)
\nonumber\\
U(1)^3            &:& \frac{4 N(N+1)^3}{F} -N(2N + 1)
\nonumber\\
U(1) U(1)_R^2     &:& -N(2N + 1)
\nonumber\\
U(1)^2  U(1)_R    &:& -N(2N + 1)  \nonumber
\eeq

\subsection{Integrating out flavors}
Another important check is to add a mass term for some of the $Q$'s and
see that this gives consistent results in the original and dual descriptions.
Consider adding a superpotential that gives masses to two of the quarks in
the original theory:
\beq
\de W = m Q_{2F - 1} Q_{2F}.
\eeq
In the first dual this is mapped to
\beq
\de W \to m \left( (M_0)_{2F - 1, 2F} -(M_0)_{2F, 2F - 1}\right).
\eeq
The $M_0$ equation of motion then requires $\twi{Q}_{2F - 1}, \twi{Q}_{2F}$
to have vacuum expectation values, breaking $\Sp{2F + 2} \to \Sp{2F}$.
Also, some of the components of the other fields become massive, and we
find that the low-energy theory is precisely the first dual description of
a theory with $F - 1$ flavors, as required for consistency.
We see the familiar pattern that integrating out a flavor in the original
theory corresponds to spontaneously breaking the gauge group in the dual
description.

In the second dual, the discussion is somewhat more complicated.
The mass term in the original theory again maps to $m (M_0)_{2F - 1, 2F}$.
The equations of motion require vacuum expectation values for $\dtwi{x}_1$
and $\twi{m}_1$, which break $\Sp{2F + 2} \to \Sp{2F}$ and
$\SO{4F + 4} \to \SO{4F}$.
Again, some of the components become massive, and one can show that the
resulting low-energy theory is precisely the second dual for $F - 1$
flavors.

Another potential check on our dual descriptions would be to add a mass term
for the
adjoint field in the superpotential.  However, in our duals this yields a
theory that is strongly coupled for all values of $N$ and $F$, so it does not
provide an additional consistency check.

\subsection{Moduli space}
We can also check that the moduli spaces are the same in the original
and the dual descriptions.
For example, consider a direction in moduli space corresponding to
$\avg{A} \ne 0$, $\avg{Q} = 0$.
Imposing the $D$-flatness condition, the simplest possibility is
\beq
\avg{A} = \pmatrix{a \si_3 & & & \cr & 0 & & \cr & & \ddots & \cr & & & 0 \cr}.
\eeq
This breaks the gauge symmetry $\Sp{2N} \to \Sp{2N - 2} \times \U{1}$,
and the massless fields transform under the unbroken gauge symmetry as
$2F$ fundamentals, an adjoint, and $4F + 1$ singlets.
The fundamentals and adjoint are neutral under the \U{1} gauge symmetry,
so the theory breaks up into three decoupled sectors in the far
infrared:
the first is identical to the original theory with $N$ reduced by one (and
$F$ unchanged);
the second has a \U{1} gauge group with $2F$ pairs of oppositely charged
matter fields;
the third is a single free chiral superfield.
There is no superpotential in this description.

In the first dual description, this vacuum corresponds to $\avg{A_1} \ne 0$,
which breaks $\SO{2N + 5} \to \SO{2N + 3} \times \U{1}$.
It is easy to check that the low-energy theory again consists of three
sectors:
the first is exactly the dual description of the {\sl Sp} sector
described above, and the second and third are identical to the corresponding
sectors above.
Note that in this dual, the physics of this vacuum is described by
spontaneous breaking of the gauge group in both the original and the
dual description.

One can consider more complicated vacua where $\avg{A}$ has more non-zero
eigenvalues by iterating the analysis above.
One might worry about the fact that the field $A_1$ in the dual description
apparently has $N + 2$ degrees of freedom along the $D$-flat direction,
while the field $A$ in the original description has only $N$ degrees of
freedom.
However, it is easy to see that giving non-zero vacuum expectation
values to $N$ components of $A_1$ in the dual leads to a confining
theory which does not have any additional flat directions corresponding to
adjoint VEVs; the apparent extra flat directions of the dual have been lifted
by strong gauge dynamics.

As another example, consider a direction in moduli space corresponding to
$\avg{Q} \ne 0$, $\avg{A} = 0$ in the original description of the theory.
Imposing the $D$-flatness condition, the simplest possibility is
\beq
\avg{Q} = \pmatrix{a 1_2 & & & & \cr & 0 & & & \cr & & \ddots & & \quad \cr
& & & 0 & \cr}.
\eeq
This breaks the gauge symmetry $\Sp{2N} \to \Sp{2N - 2}$, and the
massless fields decompose under $\Sp{2N - 2}$ as an adjoint, $2F$
fundamentals, and $4F$ singlets.
There is no superpotential.

In the first dual description, this vacuum corresponds to $\avg{M_0} \ne 0$.
This breaks the flavor symmetry $\SU{2F} \to \SU{2F - 2} \times SU(2)$, and
some of the fields $\twi{Q}$ become massive.
The gauge group is not broken in this theory, but the theory is more
strongly coupled because there are fewer matter fields.
This description is not obviously dual to the one discussed in
the previous paragraph.
To see that they are equivalent, take the dual of the $\Sp{2F + 2}$ gauge
group in this description.
We then obtain a theory that is similar to the deconfined description of
the theory.
The {\sl SO} gauge group of this description is confining, and writing
the low-energy theory of the {\sl SO} mesons we recover the description
above.

We can analyze more complicated vacuum expectation values for
$Q$ by iterating the above analysis.
We therefore have a consistency check that can be written diagrammatically
as
\beq
\begin{array}{clclcll}
\hbox{original}
& \to & \hbox{deconfined}
& \to & \hbox{dual}
\\
\downarrow &&  && \downarrow
\\
\hbox{original with VEV}
& \leftarrow & \hbox{deconfined with VEV}
& \leftarrow & \hbox{dual with VEV}
\end{array}
\eeq
where the horizontal arrows denote duality or (de)confinement
transformations, and the vertical arrows denote taking VEVs corresponding
to a given direction in moduli space.

\subsection{$N = 0$}
An amusing consistency check is to consider the theory with $N = 0$.
In this case the original theory is trivial, but our dual descriptions
appear at first sight to be non-trivial.
Consider the first dual.
In this theory, the {\sl Sp} group has the right number of flavors to
confine without breaking chiral symmetry \cite{Seib2,IntPoul}.
The fields $\twi{Q}$, $\twi{x}_1$, and $\twi{p}_2$ confine into mesons
that combine with $M_0$, $A_1$, $m_1$, and $m_2$ to become massive.
This leaves an effective theory with the field $p_3$ and a meson field
$M = \twi{x}_1 \twi{p}_2$, with superpotential
\beq
W =  p_3 M^2.
\eeq
$p_3$ is a singlet under $\SO{5}$, while $M$ transforms in the
defining \rep.
An $\SO{5}$ gauge theory with one flavor confines without breaking chiral
symmetry \cite{IntSeibSO}, so the the low-energy theory can be written in
terms of the composite meson $N = M^2$.
The superpotential then gives a mass to $N$ and $p_3$, leaving a low-energy
theory with no massless degrees of freedom.
This is exactly what is required for consistency with the original
theory.

\section{Deconfining Arbitrary 2-Index Tensors}
The methods we have used can be extended to write dual descriptions
of any gauge theory with {\sl SU}, {\sl SO}, or {\sl Sp} gauge groups
containing at most 2-index tensor representations.
Note that \susc\ gauge theories with matter in 3-index tensor representations
are not asymptotically free for large $N$ (specifically, $N > 5$ for \Sp{2N},
$N > 8$ for \SO{N}, and $N > 12$ for \SU{N}).%

In this sense, these methods allow us to construct dual descriptions of
``almost all" \susc\ gauge theories with tensor representations.

As discussed in the main text, the idea is to ``deconfine'' all the 2-index
tensors by introducing new confining gauge interactions whose low-energy
dynamics is a theory of mesons.
The simplest approach is to introduce only those fields required to produce
the 2-index tensor as a bound state, but then the number of anomaly free
$U(1)$ symmetries do not match
because there is an extra anomaly cancellation constraint from the confining
gauge group.
Also, if the confining gauge group is {\sl Sp} or {\sl SU}, the mesons have
an unwanted dynamical superpotential.
These problems are solved simultaneously by adding additional fields that
are fundamentals under the confining gauge group, together with some singlets
and a superpotential to make the additional mesons massive.
The result is a gauge theory containing only fundamental representations of
all gauge groups, and one can apply known dualities to obtain dual
descriptions from this.

For an antisymmetric 2-index tensor $X^{ab}$ with $a,b = 1, \ldots, N$
transforming under some gauge group $G$, one
introduces an additional $\Sp{2N'}$ gauge group with matter fields
\beq
(x)^{a a^\prime},
\quad
(p_1)^{a^\prime j},
\quad
j = 1, \ldots, K
\eeq
for some $K$.
In addition, one introduces $\Sp{2N'}$ singlets
\beq
(p_2)_{a,j}\, ,
\quad
(p_3)_{jk}
\eeq
with superpotential couplings
\beq
\de W = x p_1 p_2 + p_1 p_1 p_3,
\eeq
where all indices are contracted in the obvious way.
The fields with their transformation properties are displayed in Table 7.

\begin{table}[htbp]
\label{decotable}
\centering
\begin{tabular}{|c|c|c|c|}\hline
field &$ G $&$\Sp{2N'}$ & $\SU{K}$
\\\hline \hline
$X$ & $\Yasymm$& $$& $ \vbr$ \\  \hline \hline
$x$ & $\Yfund$&$\Yfund$&${\bf 1} \vbr$ \\ \hline
$p_1$ & ${\bf 1}$&$\Yfund$&$\Yfund \vbr$ \\ \hline
$p_2$ & $\overline{\Yfund}$&${\bf 1}$&$\overline{\Yfund} \vbr$ \\ \hline
$p_3$ & ${\bf 1}$&${\bf 1}$&$\overline{\Yasymm} \vbr$ \\ \hline
\end{tabular}
\parbox{4in}{\caption{Field content of the deconfinement module for
an antisymmetric tensor.}}
\end{table}

If we take $N' = \frac{1}{2}(N + K) - 2$ (choosing $K$ so that $N'$ is an
integer) then this theory confines and gives rise to a low-energy theory
with of a single meson field, which can be identified as
\beq
X^{ab} = \ep_{a^\prime b^\prime} (x)^{a a^\prime} (x)^{b b^\prime}.
\eeq
Note that for $K > 1$, there is an additional global $\SU{K}$ symmetry in
the ultraviolet, but the only fields that transform
under this symmetry are
not present in the low-energy theory.
We can now write a dual description by applying the known duality for theories
with matter only in the fundamental representation to the group $G$.

Symmetric 2-index tensors are treated in detail in Section 3, and so the only
case left to discuss is an adjoint $X^a{}_b$ of $\SU{N}$.
(Adjoint representations of {\sl SO} are antisymmetric tensors.)
We ``deconfine'' the adjoint by introducing a new $\SU{N'}$ gauge group
with matter fields
\beq
(x)^{aa^\prime},
\quad
(\mybar{x})_{aa^\prime},
\quad
(p_1)^{a^\prime j},
\quad
(\mybar{p}_1)_{a^\prime j^\prime}\ .
\eeq
(Note the bar does not indicate complex conjugation.)
If we choose $N' = N + K -1$, then this theory confines and gives rise to
a low-energy effective theory consisting of composite mesons and baryons
and a non-perturbative superpotential.
To eliminate the unwanted states and the non-perturbative superpotential
we add the following fields
\beq
p_2, \quad
(p_3)_{a}^{j^\prime}, \quad
(\mybar{p}_3)^{a}_{j}, \quad
(p_4)_j^{j^\prime}, \quad
(p_5)^a, \quad
(\mybar{p}_5)_a, \quad
(p_6)^j, \quad
(\mybar{p}_6)_{j^\prime},
\eeq
and tree level superpotential
\beq
\de W & = & p_2 x \mybar{x} + p_3 x \mybar{p}_1 + \mybar{p}_3 \mybar{x} p_1
+ p_4 p_1 \mybar{p}_1
\\ \nonumber
&& + p_5 (x)^{N-1} (p_1)^K
+ \mybar{p}_5 (\mybar{x})^{N-1} (\mybar{p}_1)^K
+ p_6 (x)^N (p_1)^{K-1} + \mybar{p}_6 (\mybar{x})^N (\mybar{p}_1)^{K-1}.
\eeq
The term $p_2 x \mybar{x}$ eliminates the trace of the meson field
$(x\mybar{x})^a{}_b$ as a dynamical field at low energies.
The fields  $p_3$, $\mybar{p}_3$, $p_4$ eliminate other unwanted mesons,
and $p_5$, $\mybar{p}_5$, $p_6$, $\mybar{p}_6$ eliminate the baryons from
the low energy spectrum.
The only massless degree of freedom left is the composite field $X^a{}_b$
(with no superpotential), as desired.
We can now write a dual description by applying the duality of Seiberg to
the gauge group corresponding to the indices $a, b, \ldots\,$.
In this way, we can write dual descriptions for the \SU{N} Kutasov--Schwimmer
model (with no tree-level superpotential) similar to the ones constructed
above in the \Sp{2N} case.
The analysis of these duals procedes in direct analogy with that in the main
body of the paper, and will not be given here.

\begin{table}[htbp]
\label{sudecotable}
\centering
\begin{tabular}{|c|c|c|c|c|}\hline
field &$ \SU{N} $&$\SU{N'}$ & $\SU{K}$ & $\SU{K}^{\prime}$
\\\hline \hline
$X$ & ${\bf A}$& $$& $$& $ \vbr$ \\  \hline \hline
$x$ & $\Yfund$&$\Yfund$& ${\bf 1}$& ${\bf 1} \vbr$ \\ \hline
$\mybar{x}$ & $\overline{\Yfund}$&$\overline{\Yfund}$& ${\bf 1}$&
${\bf 1} \vbr$ \\ \hline
$p_1$ & ${\bf 1}$&$\Yfund$&$\Yfund$& ${\bf 1} \vbr$ \\ \hline
$\mybar{p}_1$ & ${\bf 1}$&$\overline{\Yfund}$& ${\bf 1}$&
$\overline{\Yfund} \vbr$ \\ \hline
$p_2$ & ${\bf 1}$&${\bf 1}$&${\bf 1}$& ${\bf 1} \vbr$ \\ \hline
$p_3$ & $\overline{\Yfund}$&${\bf 1}$&${\bf 1}$& $\Yfund \vbr$ \\ \hline
$\mybar{p}_3$ & $\Yfund$&${\bf 1}$& $\overline{\Yfund}$& ${\bf 1}  \vbr$ \\
\hline
$p_4$ & ${\bf 1}$&${\bf 1}$&$\overline{\Yfund}$& $\Yfund \vbr$ \\ \hline
$p_5$ & $\Yfund$&${\bf 1}$& ${\bf 1}$& ${\bf 1} \vbr$ \\ \hline
$\mybar{p}_5$ & $\overline{\Yfund}$&${\bf 1}$& ${\bf 1}$& ${\bf 1} \vbr$ \\
\hline
$p_6$ & ${\bf 1}$&${\bf 1}$&$\Yfund$& ${\bf 1} \vbr$ \\ \hline
$\mybar{p}_6$ & ${\bf 1}$&${\bf 1}$& ${\bf 1}$&
$\overline{\Yfund} \vbr$ \\ \hline
\end{tabular}
\parbox{4in}{\caption{Field content of the deconfinement module for
an \SU{N} adjoint.}}
\end{table}


\end{document}